# What Is Armchair-Zigzag Grain Boundary Structure in Graphene?

Zheng-Lu Li, Hai-Yuan Cao, Ji-Hui Yang, Qiang Shu, Yue-Yu Zhang, Hongjun Xiang,* and Xingao Gong*

Key Laboratory of Computational Physical Sciences (Ministry of Education), State Key Laboratory of Surface Physics, and Department of Physics, Fudan University, Shanghai 200433, P. R. China

**ABSTRACT:** We have developed a new global optimization method for the determination of interface structure based on the differential evolution algorithm. Here, we applied this method to search for the ground state atomic structures of the grain boundary between the armchair and zigzag oriented graphene. We find two new grain boundary structures with considerably lower formation energy of about 1 eV nm$^{-1}$ than that of the previously widely used structural models. These newly predicted structures show better mechanical property under external uniaxial strain, and distinguishable scanning tunneling microscope features, compared with the previous structural models. Our results provide important new information for the determination of grain boundary structures and henceforth the electronic properties of defected graphene.

## ■ INTRODUCTION

Graphene, the two dimensional (2D) material, shows great application potential[1-3] with the advancement in techniques such as chemical vapor deposition that makes the large-scale growth of graphene feasible.[4-9] In practice, however, graphene is always grown with different defects, among which grain boundary (GB) is one of the most frequently formed defects.[10-12] GBs provide numerous novel possibilities in modifying graphene such as tuning the charge distribution[12] and transport property[13]. Therefore, GBs in graphene have been the focus of numerous researches due to their great significance in science and application.[10-15]

Intensive works have been done to study the broad properties of GBs.[13-25] By analyzing the symmetry between the Brillouin zones of two sides, a theory was developed to predict the electronic transport property through GBs,[13] indicating symmetric GBs have zero transport gap, but the asymmetric GBs have finite gaps. Such predictions are confirmed by non-equilibrium Green's function (NEGF) calculations,[13,16] and provide promising potential to regulate the electric transport properties. As a result, some heterojunctions have already been designed using GBs to develop new transport devices.[17,18] The mechanical responses of under tensile stress were studied by molecular dynamics (MD) simulations.[19-22] It is found that the strength of graphene with GBs are affected by not only the density of GBs,[19] but also the local structures at GBs.[19,20] Thermal properties were examined by both NEGF[23] and MD[24] techniques, indicating the excellent thermal conductivity of GBs.[23]

Structures are the basis of theoretical investigations on materials, so are the GBs in graphene. Among all GBs in graphene, a certain type of GB between armchair and zigzag oriented graphene is particularly interesting.[13,15-18,21-25] People wonder how graphene behaves when the two different types of edges with distinct structural, electronic and magnetic properties meet. For this type of GB, many of the abovementioned works[13,15-18,21-25] adopted two similar GB structures shown in Figure 1(a) and (c), denoted as GB-I and GB-i in this work, corresponding to (7, 0)|(4, 4) and (5, 0)|(3, 3) lattice matching respectively. Both GB-I and GB-i present the same structural character that one pentagon and two heptagons gathered at one point (flyhead pattern). However, the gathering of defects is likely to increase the intrinsic stress,[22] making the structure relatively unstable with high energy. Thus, a quite serious and questionable issue is whether GB-I or GB-i is the ground state structure of the GB between the armchair and zigzag oriented graphene. Therefore, it is desirable to reinvestigate this GB structure using more sophisticated approach.

Unfortunately, the prediction of the interface structure is a very difficult task, although global optimization methods have been successfully applied to predict both two and three dimensional (3D) crystal structures.[26-33] Usually, the interface structure is much more complex than the corresponding bulk systems. To predict interface structures, much more factors must be considered, largely increasing the complexity of the task. Initial efforts towards interface structure prediction have been undertaken in some precedent works[34,35] using global optimization algorithms.

Among the numerous global optimization algorithms, differential evolution (DE) has been applied to many fields and achieved great successes.[36-40] DE is based on the idea that using the differentials of randomly selected solution candidates to mutate the existing ones and the generated candidates are accepted only if they have improvements (a greedy strategy). DE is strongly believed to have competitive performance in structure searching because previous studies showed DE outperforms many other algorithms with the tested functions and problems.[36,38,40]

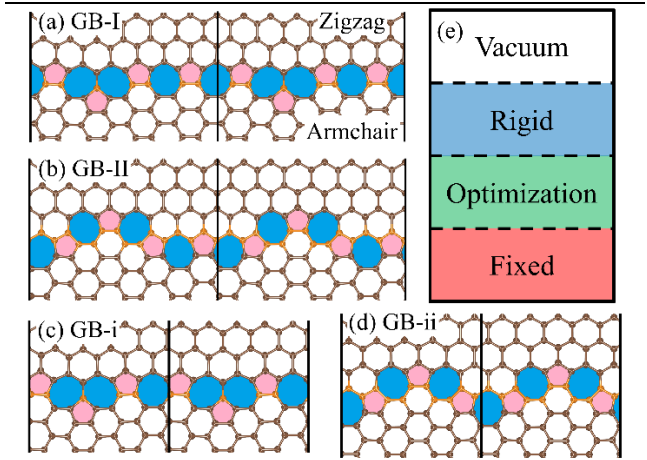

**Figure 1.** (a), (c) The previously widely used structures of GB between armchair and zigzag oriented graphene with the gathering of one pentagon and two heptagons, denoted as (a) GB-I in (7, 0)|(4, 4) and (c) GB-i in (5, 0)|(3, 3). (b), (d) The presently found GB structures with an armchair-like shape, denoted as (b) GB-II in (7, 0)|(4, 4) and (d) GB-ii in (5, 0)|(3, 3). Atoms in the optimization layer are marked using gold color. (e) A schematic illustration of the slab model used in our interface structure prediction.

In this paper, to generally solve the global optimization of the 2D (the GBs in graphene could be viewed as 2D interfaces) and 3D interfaces, we developed a method based on DE algorithm to theoretically predict the interface structures. The performance of our method turns out to be very efficient in searching interface structures. We have found two new GB structures between the armchair and zigzag oriented graphene in two lattice matching cases respectively. In (7, 0)|(4, 4) [(5, 0)|(3, 3)] GB, the newly found structure, denoted as GB-II (GB-ii), has a considerably lower of 1.07 eV nm$^{-1}$ (1.04 eV nm$^{-1}$) than the widely accepted structure GB-I (GB-i). The new GB-II and GB-ii structures [Figure 1(b) and (d)] are also quite similar, consisting of pentagons and heptagons with no gathering but an armchair-like chain. Among all structures we considered in this work, the newly found GB-II has the lowest formation energy, better mechanical properties under uniaxial strain, and could be distinguished by STM with other structures. The new structures found in this paper provide different structural basis for the further investigations on GBs in graphene.

## ■ METHODS

**DE Based Global Optimization Method for Interface Structure Prediction.** In our approach, we use the slab model to simulate the interface system. As shown in Figure 1(e), we set up the structure with different layers stacking along the $c$ axis, which is set perpendicular to the $ab$ plane, and the periodic boundary conditions (PBC) are applied on $a$ and $b$ directions. At the bottom is the fixed layer, which means the atoms in this layer is fixed during the whole searching process to simulate the bulk. The optimization layer is in the middle, corresponding to the interfacial region. On the top is the rigid layer, in which the atoms always keep the relative coordinates constant, but could translate as a whole rigid body. During the simulation, besides the degrees of freedom (DOF) of the atomic positions (2 and 3 for the 2D and 3D cases, respectively) in the optimization layer, we also allow three more DOF: the height of the optimization layer and the translation along $a$ and $b$ directions for the rigid layer. This operation would lead to more reliable results by avoiding the constraints brought by the initial setup. The dimension of the problem in the 3D case is thus $3N_{opt} + 3$, where $N_{opt}$ is the number of atoms in the optimization layer.

DE is a global optimization algorithm designed to search the multidimensional continuous spaces to best fit to the designated evaluation functions.[36-40] In the basic DE algorithm, each solution candidate is treated as a vector in the $D$-dimensional phase space, and involves in three steps: mutation, crossover and selection. The mutation operation generates a mutant vector **v** for the $i$th target vector **x** in the population as follow:

$$\mathbf{v}_{i,G+1} = \mathbf{x}_{r1,G} + F(\mathbf{x}_{r2,G} - \mathbf{x}_{r3,G}) \quad (1)$$

where $G$ denotes the generation, $r1$, $r2$ and $r3$ are random indexes in the population which are mutually different from $i$, $F$ is a parameter that controls the effect of differential vector. Crossover step creates the trail vector $\mathbf{u}_{i,G+1} = (u_{1i,G+1}, u_{2i,G+1}, \dots, u_{Di,G+1})$ according to the following scheme:

$$u_{ji,G+1} = \begin{cases} v_{ji,G+1}, \text{if } r(j) \leq CR \text{ or } j = rn(i) \\ x_{ji,G}, \text{if } r(j) > CR \text{ and } j \neq rn(i) \end{cases} \quad (2)$$

where $r(j) \in [0,1]$ is the $j$th uniformly generated random number, $rn(i)$ represents a randomly chosen index of dimension to ensure the $i$th target vector gets at least one element from the mutant vector, and $CR \in [0,1]$ is the crossover probability. Selection in DE simply takes the greedy principle to accept the trail vector only if it is better than the previous corresponding target vector. Two important parameters $F$ and $CR$ in DE control the general behaviors of the algorithms and in this work we choose $F = 0.5$ and $CR = 0.9$. To optimize interface structures using DE, each potential structure corresponds to one target vector with $D = 3N_{opt} + 3$.

The use of symmetry constraints[31,32] is suggested to significantly improve the performance of global optimization.

However, usually the interfacial part has rather low symmetry since the bulk structures on the two sides are always asymmetric. To achieve the high efficiency of global optimization based on DE, we propose another strategy. In DE, operations are applied dimension by dimension. However, it might happen that the vectors corresponding to two interface structures are misaligned, which results in unrealistic high energy structure by the DE operations. Thus, we take a sorting strategy for the atoms in the optimization layer according to their positions, so that structures would not be distorted too much or even destroyed by DE. With this additional sorting step, the efficiency of the global optimization by DE could be kept high.

After all structures are generated by DE in each generation, all the $3N_{opt} + 3$ DOF of every structure will be relaxed to its local minimum, using either empirical potentials or first-principles calculations. In this work, because of the 2D nature of graphene, the DOF is reduced to $D = 2N_{opt} + 2$. The local optimization is a common routine in structure prediction, aiming to effectively reduce the search space and to get total energy of each structure.

**Empirical Potentials Calculations.** Our empirical potentials calculations are performed using LAMMPS[41] with the widely used AIREBO potential[42] for graphene systems. For local optimization in DE searching, we minimize the total energy of every structure using conjugate gradient method. For the molecular dynamics (MD) simulations which are used to get the mechanical property of the structure finally predicted by DE, we use the following method:[20] MD simulations are performed using *NVE* ensemble, i.e., atom number, volume and energy are constant, the cutoff radius of C-C bond $r_{CC}$ is set 1.92 Å, the graphene sheets are applied with uniaxial strain at a rate of $10^{-9}$ s$^{-1}$, and Virial stresses are calculated.

**First-principles Calculations.** The first-principles calculations based on DFT are performed using VASP[44] with the projected augmented wave method.[45,46] We use first-principles calculations to ensure the final results from DE and to study the electronic properties. For such purposes, we use the local density approximation to describe the exchange-correlation potential in the DFT calculations. Structures are relaxed until the atomic forces are less than 0.01 eV/Å and total energies are converged to $10^{-6}$ eV with the cutoff energy for plane-wave basis wave functions set to 400 eV.

**Searching Criterion.** We intend to search the structure of the GB between the armchair and zigzag oriented graphene, as shown in Figure 1. Since we use the slab model to simulate the interface structure with one dimensional PBC, the structure we used is actually carbon nanoribbon with GB parallel to the edges. The two edges are passivated by hydrogen atoms, and are about 15 Å away from the GB. For each case of lattice matching, we keep the fixed and rigid layers the same, and perform a series of searching with different numbers of atoms in the optimization layer. We adopt a simple searching criterion to minimize the relative formation energy defined below:

$$\Delta E_{form}(N_{opt}) = [E_{tot}(N_{opt}) - E_{tot}(GB\text{-}I) - (N_{opt} - 7) \times \mu_C]/L \quad (3)$$

where $\Delta E_{form}$ is the relative formation energy, $E_{tot}(N_{opt})$ the total energy of the structure with a certain $N_{opt}$, and $\mu_C$ the chemical potential of one carbon atom taken from pristine graphene. In fact, we are using GB-I as the zero point in the comparisons, i.e., $\Delta E_{form}(GB\text{-}I) = 0$, because $N_{opt} = 7$ for GB-I (see Table 1). This equation is used in searching structures specifically in (7, 0)|(4, 4) GB, with a certain periodicity $L$ along GB. For (5, 0)|(3, 3) GB, Eq. (3) could be rewritten by substituting GB-i for GB-I and $(N_{opt} - 5)$ for $(N_{opt} - 7)$, because $N_{opt} = 5$ for GB-i (see Table 1).

To get the absolute graphene GB formation energy so that we could compare different structures in a more direct manner, we construct the periodic structures made of only carbon atoms by applying inversion symmetry to the slabs we used in the global optimization. The absolute formation energy could be written as:[21]

$$E_{form} = (E_{tot} - N \times \mu_C)/2L \quad (4)$$

where $E_{tot}$ denotes the total energy of the periodic cell containing two equivalent GBs with the length of GB $L$, and $N$ the total number of atoms in the cell.

## RESULTS AND DISCUSSIONS

**Predicting Grain Boundary Structure.** As a benchmark, we firstly used our algorithm to search a well-known symmetric GB and found the stable structure quite easily (Supporting Information). For the (7, 0)|(4, 4) GB, we scan a large range of $N_{opt}$ and plot our results in the upper panel of Figure 2. When $N_{opt} = 15$, a new GB structure is found, shown as GB-II in Figure 1(b). After full structural relaxation by DFT, we obtain $E_{form}$ (GB-I) = 4.29 eV nm$^{-1}$ and $E_{form}$ (GB-II) = 3.22 eV nm$^{-1}$ (see Table 1), indicating that the formation energy of GB-II is 1.07 eV nm$^{-1}$ (a considerably large formation energy difference) lower than GB-I. The reason for this is quite simple. GB-I structure possesses a gathering of one pentagon and two heptagons, while GB-II is just a clean pentagon-heptagon chain with armchair-like shape. Since the gathering of the pentagon and heptagon defects in graphene usually leads to higher energy, GB-II could effectively reduce the formation energy comparing with GB-I. Interestingly, when $N_{opt} = 14$, we found a structure [see upper panel of Figure 2] that has slightly lower formation energy than GB-I by empirical potential calculations. This structure has a hole in the GB and actually if one additional carbon atom is added to the middle of the hole, it would become GB-II.

**Table 1. Formation energies of four GBs. $N_{opt}$ is the number of atoms in the optimization layer.**

| GBs | GB-I | GB-II | GB-i | GB-ii |
|---|---|---|---|---|
| $N_{opt}$ | 7 | 15 | 5 | 11 |
| $E_{form}$ (eV nm$^{-1}$) | 4.29 | 3.22 | 5.45 | 4.41 |

For (5, 0)|(3, 3) GB, we also find a new structure GB-ii, which has a considerably lower formation energy of 1.04 eV nm$^{-1}$ than GB-i (see Table 1). Actually, GB-I and GB-i are similar, and GB-II is similar to GB-ii as well (see Figure 1): comparing with GB-I, GB-i lacks one pentagon-heptagon pair; if we take one pentagon-heptagon pair away from GB-II, it naturally becomes GB-ii. However, (5, 0)|(3, 3) GB has a larger lattice mismatch than (7, 0)|(4, 4) GB (3.8% to 1.0%). According to Eq. (4), when calculating the absolute formation energy, with two equivalent GBs in one unit cell, the farther the two GBs separate, (5, 0)|(3, 3) GB would have larger formation energy than (7, 0)|(4, 4) GB, because the graphene part between GBs would get higher energy due to the strain effect. To avoid ambiguity, we address here that for all periodic structures in this work, we have two equivalent GBs in one unit cell separating about 30 Å with each other. Because (7, 0)|(4, 4) GB has a smaller lattice mismatch and lower formation energy (see Table 1), practically it is more preferable experimentally, and we would mainly focus on this type of GB. In all four structures considered in this work, GB-II has the lowest formation energy (see Table 1).

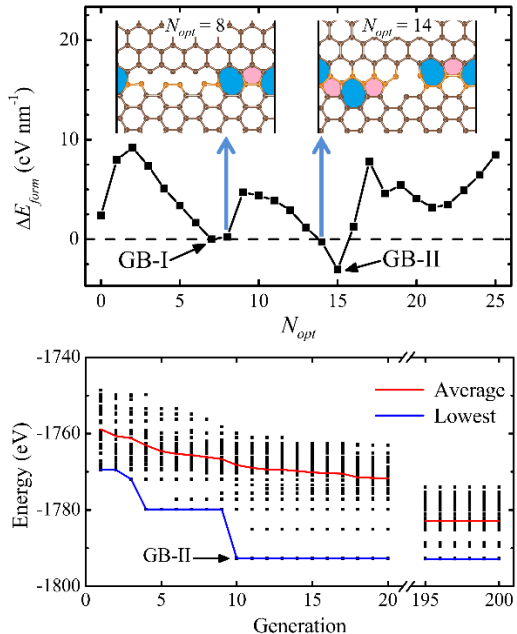

**Figure 2.** The upper panel: the relative formation energy $\Delta E_{form}$ to GB-I by empirical potential for each $N_{opt}$. GB-II structure is found to have the lowest formation energy. Structures with $N_{opt}$ = 8 and 14 are shown in insets. The lower panel: performance of our method based on DE algorithm. Black dots represent the total energy evaluated for the structures in each generation. GB-II is found in 10 generations, indicating high efficiency of our method.

Before further investigation into the properties of the newly found GB-II, we first demonstrate the global optimization efficiency of our algorithm for (7, 0)|(4, 4) GB. In our DE simulations, we set the size of population to 30, and the maximum generation number 50. To check the reliability and efficiency of our methods, we have performed ten separate independent DE simulations $N_{opt}$ = 15 and all found GB-II structure. The smallest number of generation of finding GB-II is 4, the largest 23, and the average 12.2, evidencing highly efficient performance in this 32-dimensional optimization problem. The lower panel of Figure 2 shows the history of one typical search for the $N_{opt}$ = 15 case, with the maximum generation number set 200 for the purpose of analysis. In this case, the program finds GB-II in 10 generations. Note that the average energy of all structures keeps decreasing during the simulation owing to the greedy selection, indicating good convergence of our method based on DE.

**Mechanical Property.** Since a counterintuitive conclusion that graphene becomes stronger with higher density of defects in more tilted GBs was reported,[19] the strength of graphene with the existence of GBs attracts much research attention. Further study indicates the strength of graphene also depends on the arrangements of defects.[20] Both works pointed out the importance of detailed atomic stress of the critical bonds in GBs that could be decisive in the failure behavior of graphene under strain. Considering that our newly found GB-II with a clean pentagon-heptagon chain has much lower formation energy than the defects-gathered GB-I, it is expected that GB-II possesses better mechanical property.

We expand the periodic structures to supercells about 100 nm in length and 30 nm in width. We apply the uniaxial strain along the directions either perpendicular or parallel to the orientation of GBs, and get the corresponding stress components of $\sigma_{zz}$ and $\sigma_{xx}$, respectively, as shown in Figure 3(a). In both directions, GB-II outperforms GB-I on the strain at failure and the final stress, showing a better mechanical response to the external engineer strain (perpendicular to GB: 6% larger strain and 2% larger stress; parallel to GB: 5% larger strain and 9% larger stress). We show in Figure 3(b) the bond length distribution in GBs, i.e., those form the pentagons and heptagons. Clearly, bonds of GB-II are much closer to the bond length in pristine graphene, while bonds of GB-I appear more dispersive in length, indicating larger structural distortion and thus weaker mechanical strength of GB-I compared with GB-II. To understand the failure behavior of both structures from a microscopic perspective, we plot the bond breaking of GB-I and GB-II in Figure 3(c) and (d) with strain applied perpendicular to GBs, since this component of stress directly reflects the GB normal strength [43]. We find two common features in the bond-breaking process of both GB-I and GB-II. First, bond-breaking begins in GB. In GB-I, it is the shared edge of two heptagons at the gathering of three defects that first breaks, and in GB-II, the bond shared by one pentagon and one heptagon breaks first; Second, bonds that are (almost) parallel to the strain direction are the first to break, which is easy to understand because the increase in the bond length of such bonds is the largest.

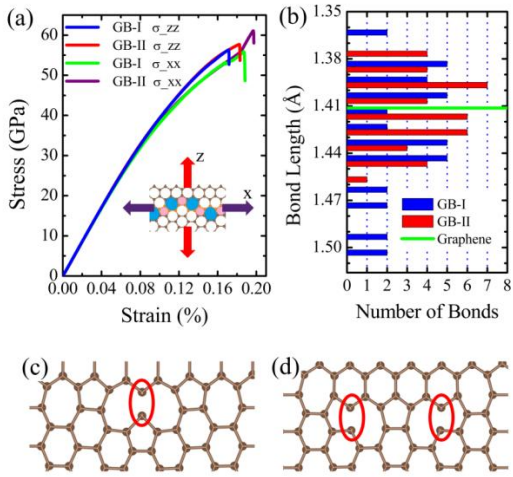

**Figure 3.** (a) The stress of GB-I and GB-II with strain applied perpendicular and parallel to the GBs. (b) Statistics of number of bonds of GBs. Pristine graphene has only one bond length at 1.41 Å (DFT result). Bonds in GB-I are from 1.36 Å to 1.51 Å, while bond lengths in GB-II are more close to that of pristine graphene. (c), (d) Schematic showing of bond breaking of GB-I and GB-II, respectively, with strain perpendicular to the GBs.

**Electronic Structure.** Electronic structure is one of the major properties of graphene, with both scientific and technological importance. To help comparing the electronic structures of GB-I and GB-II, we perform first-principles calculations using the periodic structures (not supercell but unit cell) that we mentioned above.

Along the GB direction in the reciprocal space, both structures are metallic, as indicated by the band structures in Figure 4(a) and (b). Difference between GB-I and GB-II appears at the second band above the Fermi level at the Γ point: the marked band [pointed by arrow in Figure 4(a)] of GB-I is less dispersive than that [pointed by arrow in Figure 4(b)] of GB-II, indicating the marked band is more localized in GB-I. Partial charge density distribution (shown as insets in Figure 4) of the marked band at Γ point in two structures confirm that the states are localized at the GBs, and this state in GB-II is more delocalized in real space. This may due to the certain topology of the two GB structures. The states of the marked bands are mainly localized to the gathering point of the one pentagon and two heptagons in GB-I, whereas are relatively evenly distributed along the pentagon-heptagon chain in GB-II. The low effective electron mass of the GB related band in GB-II may suggest a better transport properties along the GB when electrons are doped by external gate voltage.

Experimentally, STM has been widely used to probe the structure of GBs in graphene.[9] According to the Tersoff-Hamann approximation[46] that the tunneling current is proportional to the local density of states of the surface, we simulate the STM images by integrating the charge density of the occupied (unoccupied) states within the range of 1.5 eV below (above) the Fermi level.[47] Figure 4(c) – (f) are the simulated STM images for GB-I and GB-II, showing the electron states right below the Fermi level are localized to the pentagons, while those right above the Fermi level are localized to the heptagons. According to our simulations, the two structures can be distinguished by STM experiment. In GB-I, the occupied states show the separate pentagon points to the dent formed by the other three pentagons [Figure 4(c)] and the unoccupied states show explicitly two heptagons touching [Figure 4(e)]. However in GB-II, the separate pentagon points away to the dent [Figure 4(d)] and all heptagons are separate [Figure 4(f)]. Thus, STM becomes an effective technique to verify our newly found GB-II structure experimentally.

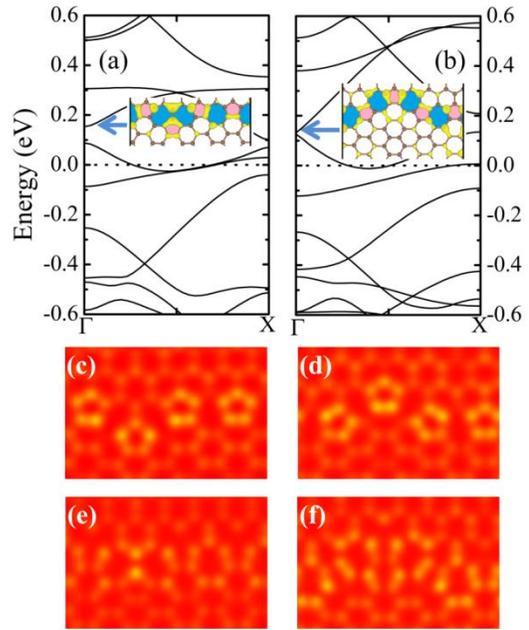

**Figure 4.** (a), (b) Band structures calculated for GB-I and GB-II, respectively. The blue arrows point out the bands of the GB (defects) states, showing different properties in two structures. The insets show partial charge density distribution (in yellow) of the marked band at Γ point of two structures, indicating that of GB-II is more delocalized. (c), (d) Simulated STM images for occupied states of GB-I and GB-II, respectively. (e), (f) Simulated STM images for unoccupied states of GB-I and GB-II, respectively. The two structures are distinguishable by STM.

Usually for a normal zigzag carbon nanoribbon, the two edges are antiferromagnetically (AFM) coupled, but it is proved by first-principles calculations that the ferromagnetism of the zigzag edge is sustained with the other edge armchair shape, using a strained GB-I structure.[25] To see whether the ferromagnetism of the zigzag edge is still stable in our new GB structure, we performed spin-polarized calculations. We narrow the slabs of GB-I and GB-II in consistent with the previous calculations[25] (Sup-porting Information). We found the ferromagnetism of the zigzag edge of two structures are both sustained. We define $\Delta\varepsilon = (E_{FM} - E_{AFM})/n_z$ to represent the energy difference of FM and AFM states of the zigzag edge with $n_z$ the number of carbon atoms at this edge. We found $\Delta\varepsilon(GB\text{-}I) = -0.011$ eV and $\Delta\varepsilon(GB\text{-}II) = -0.012$ eV, which indicates that FM is even more enhanced in GB-II. As for the FM states, the magnetic moment averaged to each carbon atom at the zigzag edge is 0.197 $\mu_B$ and 0.210 $\mu_B$ for GB-I and GB-II, respectively. Taken together, GB-II structure could better stabilize the FM state of the zigzag edge comparing with GB-I, and could also increase the FM moment on the edge.

# ■ CONCLUSION

To predict the structure of interface, we develop a global optimization method using DE algorithm. We apply our method to searching the structure of GB between the armchair and zigzag oriented graphene, and find the new structure GB-II (GB-ii) that has a 1.07 eV nm$^{-1}$ (1.04 eV nm$^{-1}$) lower formation energy than previously widely used GB-I (GB-i) in (7, 0)|(4, 4) GB [(5, 0)|(3, 3) GB]. Our new method also shows great efficiency and convergence for this multidimensional problem. The newly found GB-II has the lowest formation energy among the four structures. MD simulations show that GB-II has a better mechanical property under uniaxial strain. STM image simulations show that GB-I and GB-II can be distinguished by the STM technique. Our study provides new insight on the structures and properties of GBs in graphene.

# ■ ASSOCIATED CONTENT

**Supporting Information**. Benchmark of our program, structures used in calculating the magnetism at the edge. This material is available free of charge via the Internet at http://pubs.acs.org.

# ■ AUTHOR INFORMATION

### Corresponding Author
* hxiang@fudan.edu.cn
* xggong@fudan.edu.cn

# ■ ACKNOWLEDGMENT

The authors thank M. Ji, X. Gu, S. Chen and Z. Guo for insightful discussions. The work was partially supported by the Special Funds for Major State Basic Research, National Science Foundation of China (NSFC), Program for Professor of Special Appointment (Eastern Scholar), and Foundation for the Author of National Excellent Doctoral Dissertation of China.

# ■ REFERENCES

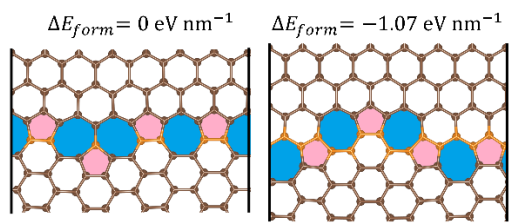

TOC